\documentclass[prb,twocolumn]{revtex4}
\usepackage{graphicx}
\newcommand{\bbb}{Bi$_{2}$Sr$_{2}$Ca$_{2}$Cu$_{3}$O$_{10}$}
\newcommand{\bb}{Bi$_{2}$Sr$_{2}$CaCu$_{2}$O$_{8}$}

\begin{document}
\title{In-plane optical spectral weight transfer in optimally doped Bi$_{2}$Sr$_{2}$Ca$_{2}$Cu$_{3}$O$_{10}$}
\author{F. Carbone, A. B. Kuzmenko,  H. J. A. Molegraaf,
E. van Heumen, E. Giannini and D. van der Marel} \affiliation{
DPMC, University of Geneva, 24, Quai Ernest-Ansermet, Geneva 4,
Switzerland\\}

\date{\today}

\begin{abstract}

We examine the redistribution of the in-plane optical spectral
weight in the normal and superconducting state in tri-layer \bbb\
(Bi2223) near optimal doping ($T_c$ = 110 K) on a single crystal
via infrared reflectivity and spectroscopic ellipsometry. We
report the temperature dependence of the low-frequency integrated
spectral weight $W(\Omega_c)$ for different values of the cutoff
energy $\Omega_c$. Two different model-independent analyses
consistently show that for $\Omega_c$ = 1 eV, which is below the
charge transfer gap, $W(\Omega_c)$ increases below $T_c$, implying
the lowering of the kinetic energy of the holes. This is opposite
to the BCS scenario, but it follows the same trend observed in the
bi-layer compound \bb\ (Bi2212). The size of this effect is larger
in Bi2223 than in Bi2212, approximately scaling with the critical
temperature. In the normal state, the temperature dependence of
$W(\Omega_c)$ is close to $T^2$ up to 300 K.

\end{abstract}

\maketitle

\section{INTRODUCTION}

The discussion as to what extent the superconductivity in
high-$T_c$ cuprates can be understood as a BCS-like
pairing-instability of a standard Fermi-liquid is open for almost
two decades. The dominating role of the Coulomb interaction in the
formation of the non-Fermi-liquid electronic state in the
underdoped region of the phase diagram questions the ability of
the BCS theory, where these interactions are treated only as a
perturbation, to universally explain the mechanism of
superconductivity in the high-$T_c$ cuprates. The pseudogap
\cite{WalstedtScience90,NormanNature93,PuchkovJPCM96}, and the
d-wave symmetry of the superconducting gap \cite{HarlingenRMP95}
are difficult to understand in terms of the standard BCS mechanism
based on the electron-phonon interaction. Another indication for
an unconventional pairing mechanism came from the experimental
temperature dependence of the low energy spectral weight of
optimally doped and underdoped
Bi2212\cite{MolegraafScience02,SantanderEPL03}, which behaves
opposite to the prediction from BCS theory\cite{MarelKluwer03}. On
the other hand, it has been shown that in the overdoped region of
the phase diagram the system has a more conventional Fermi-liquid
behavior\cite{ProustPRL02,YusofPRL02,JunodPC99,DeutscherPRB05}. In
a conventional superconductor, the internal energy is lowered as
the net result of a decrease of the charge carrier interaction
energy and a smaller increase of their kinetic energy. Several
alternative models have been proposed for the pairing mechanism of
high $T_c$ cuprates
\cite{AndersonScience87,AndersonJPCM87,HirschPC92}, many of them
predict a decrease of kinetic energy in the superconducting state
\cite{AndersonScience87,AndersonJPCM87,HirschPC92,EcklPRB03,WrobelJPCM03}.
This issue can be addressed by optical
techniques\cite{MolegraafScience02,SantanderEPL03}, taking
advantage of the relation between the intraband spectral weight
and the energy momentum dispersion of the conduction electrons
\cite{MaldaguePRB77}
\begin{equation}\label{equation1}
    W(\Omega_c) \equiv {\int}^{\Omega_c}_0 \sigma_1(\omega)d\omega =
     \frac{\pi e^2 a^2}{2\hbar^2V}<-\hat{T}>,
\end{equation}
\noindent where $\sigma_{1}(\omega)$ is the real part of the
optical conductivity, $\Omega_c$ is a cutoff frequency, $a$ is the
in-plane lattice constant, $V$ is the volume of the unit cell, and
$\hat{T}\equiv -a^{-2}\sum_k \hat{n}_k
\partial^2\epsilon_k/\partial k^2$. In the superconducting state, the
integration must include the $\delta$-peak at zero frequency due
to the condensate. In the nearest neighbor tight-binding
approximation $\hat{T}$ is exactly the kinetic energy of the
conduction band electrons, but even in the next nearest neighbor
model with $t'/t\sim-0.3$ (where $t$ and $t'$ are the nearest
neighbor and the next nearest neighbor hopping parameters
respectively) the kinetic energy and $<-\hat{T}>$ are
approximately equal and they follow the same trends as a function
of temperature \cite{MarelKluwer03}. In this case the lowering of
$W(\Omega_c)$ implies an increase of the electronic kinetic energy
and vice-versa. The value of $\Omega_c$ has to be chosen as to cut
off the region of the interband transitions. In the presence of
strong electron correlations the intraband energy region becomes
very broad as it includes the high-frequency peaks due to the
electronic transitions leading to the double occupancy. Recently
Wrobel {\em et al.} \cite{WrobelJPCM03} pointed out that
$W(\Omega_c)$ is representative of the kinetic energy within the
t-J model, if $\Omega_c$ is chosen between the values of exchange
integral $J\sim$ 0.1 eV and hopping $t\sim$ 0.4 eV, while it
corresponds to the kinetic energy of the Hubbard model when the
cutoff energy is above $U\sim$ 2 eV.

In this paper we experimentally address these issues by analysing
infrared and optical spectra of the tri-layer compound Bi2223.
This material has an advantage of a very high transition
temperature (about 110 K), resulting in a large
superconductivity-induced change of the optical conductivity, and
a high-quality cleaved optical surface. Furthermore, it belongs to
the same homologic family as the much more extensively studied
bi-layer Bi2212, making it suitable for a comparative study of the
spectral weight transfer in materials with different numbers of
CuO$_{2}$ planes per unit cell. Given the theoretical implications
related to the choice of the cut-off frequency we present the
temperature dependence of $W(\Omega_c)$ for different relevant
values of $\Omega_c$.

While the physical meaning of the temperature dependent
W($\Omega_c$) is a matter of theoretical interpretations, it can
be, in fact, experimentally determined without model assumptions.
The latter is non-trivial, since the integration in
Eq.\ref{equation1} requires, at first glance, the knowledge of
$\sigma_{1}(\omega)$ down to zero frequency and a separate
determination of the superfluid density. Fortunately, an
additional information about the real part of the dielectric
function $\epsilon_1(\omega)$, which is independently obtained
from ellipsometry as well as reflectivity measurement (since
reflectivity depends on both $\epsilon_1$ and $\epsilon_2$),
allows one to determine accurately $W(\Omega_c)$ and its
temperature dependence without low-frequency data extrapolations.
Although this point was emphasized in our previous
publications\cite{MolegraafScience02,KuzmenkoPRB05}, we reiterate
in this article the details of the experimental determination of
$W(\Omega_c)$ due to its topmost experimental importance. Two
different numerical approaches were used which gave us consistent
results: (i) the calculation of $W(\Omega_c)$ at each temperature
and (ii) the temperature-modulation analysis of superconductivity
induced changes of the optical properties\cite{KuzmenkoPRB05}.

\section{EXPERIMENT AND RESULTS}

\begin{figure}[ht]
   \centerline{\includegraphics[width=8.5cm,clip=true]{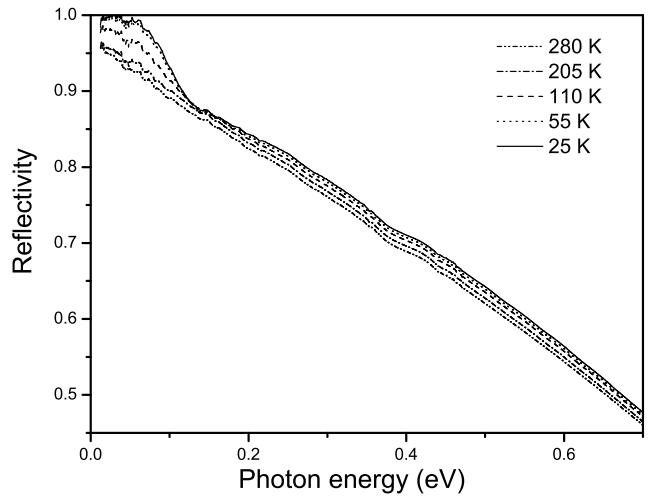}}
   \caption{In-plane reflectivity spectra of optimally doped Bi2223 for selected temperatures.}
   \label{refl}
\end{figure}

\begin{figure}[ht]
   \centerline{\includegraphics[width=8.5cm,clip=true]{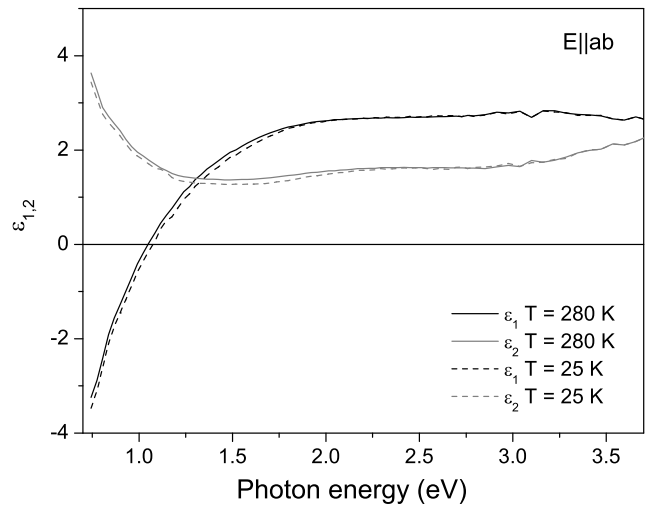}}
   \caption{Bi2223 ab-plane dielectric function at selected temperature.}
   \label{ell}
\end{figure}

Two large single crystals of \bbb\ with $T_c$ = 110 K and
transition width $\Delta T_c$ $\sim$ 1 K were prepared as
described in Ref. \onlinecite{GianniniSST04}. The samples had
dimensions ($a\times b \times c$) of $4\times1.5\times 0.2$
mm$^{3}$ and of $3\times0.8\times 3$ mm$^{3}$ respectively. The
first crystal has been used to measure the in-plane optical
properties and was cleaved within minutes before being inserted
into the cryostat. We measured the normal-incidence reflectivity
from 100 and 7000 cm$^{-1}$ (12.5 meV - 0.87 eV) using a Fourier
transform spectrometer, evaporating gold {\em in situ} on the
crystal surface as a reference. The reflectivity curves for
selected temperatures are displayed in Fig. \ref{refl}. The
ellipsometric measurements were made on the same sample surface in
the frequency range between 6000 and 36000 cm$^{-1}$ (0.75 - 4.5
eV) at an angle of incidence of 74$^\circ$. The ellipsometrically
measured pseudo-dielectric function was numerically corrected for
the admixture of the c-axis component which provided the true
ab-plane dielectric function, whose real and imaginary parts are
shown in Fig. \ref{ell}. The c-axis dielectric function that is
required for this correction was measured independently on the
ac-oriented surface of the second crystal as described in the
Appendix. It is shown in Fig. \ref{casp}.

The superconductivity induced changes of the optical properties at
photon energies above the superconducting gap (in the mid-infrared
and higher frequencies), are rather small but their reliable
detection is crucial to determine to correct sign and magnitude of
the spectral weight transfer. We used home-made optical cryostats,
whose special design preserves the sample alignment during thermal
cycling. In the visible - ultraviolet (UV) region, in order to
avoid spurious temperature dependencies of the optical constants
due to adsorbed gases at the sample surface, an ultra high vacuum
cryostat was used, operating at a pressure in the $10^{-10}$ mbar
range. All data were acquired in the regime of continuous
temperature scans at a rate of about 1 K/minute between 20 K and
300 K with a resolution of 1 K. The signal to noise ratio of the
temperature dependent reflectivity in the mid-infrared is about
2000.

\begin{figure}[ht]
   \centerline{\includegraphics[width=9cm,clip=true]{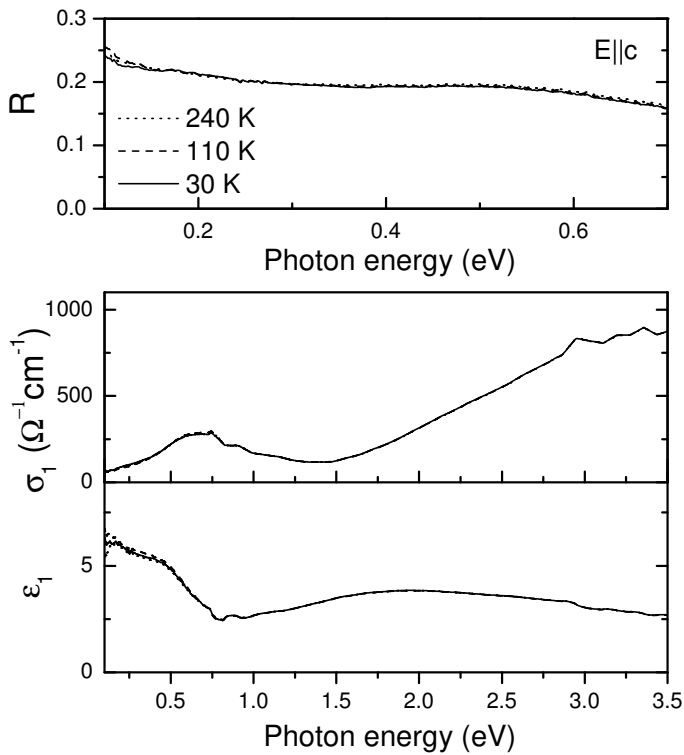}}
     \caption{c-axis optical spectra of Bi2223 at selected temperatures. Top panel: normal incidence
     reflectivity, middle panel: $\sigma_{1}(\omega)$, bottom panel: $\epsilon_{1}(\omega)$.}
\label{casp}
 \end{figure}

In order to obtain the optical conductivity $\sigma_{1}(\omega)$
in the infrared region we used a variational routine
\cite{KuzmenkoRSI05} yielding the Kramers-Kronig consistent
dielectric function which reproduces all the fine details of the
infrared reflectivity spectra while simultaneously fitting to the
ellipsometrically measured complex dielectric function in the
visible and UV-range. In contrast to the "conventional" KK
reflectivity transformation this procedure anchors the phase of
the complex reflectivity to the one at high energies measured
directly with ellipsometry\cite{BozovicPRB90}.

\begin{figure}[ht]
   \centerline{\includegraphics[width=8.5cm,clip=true]{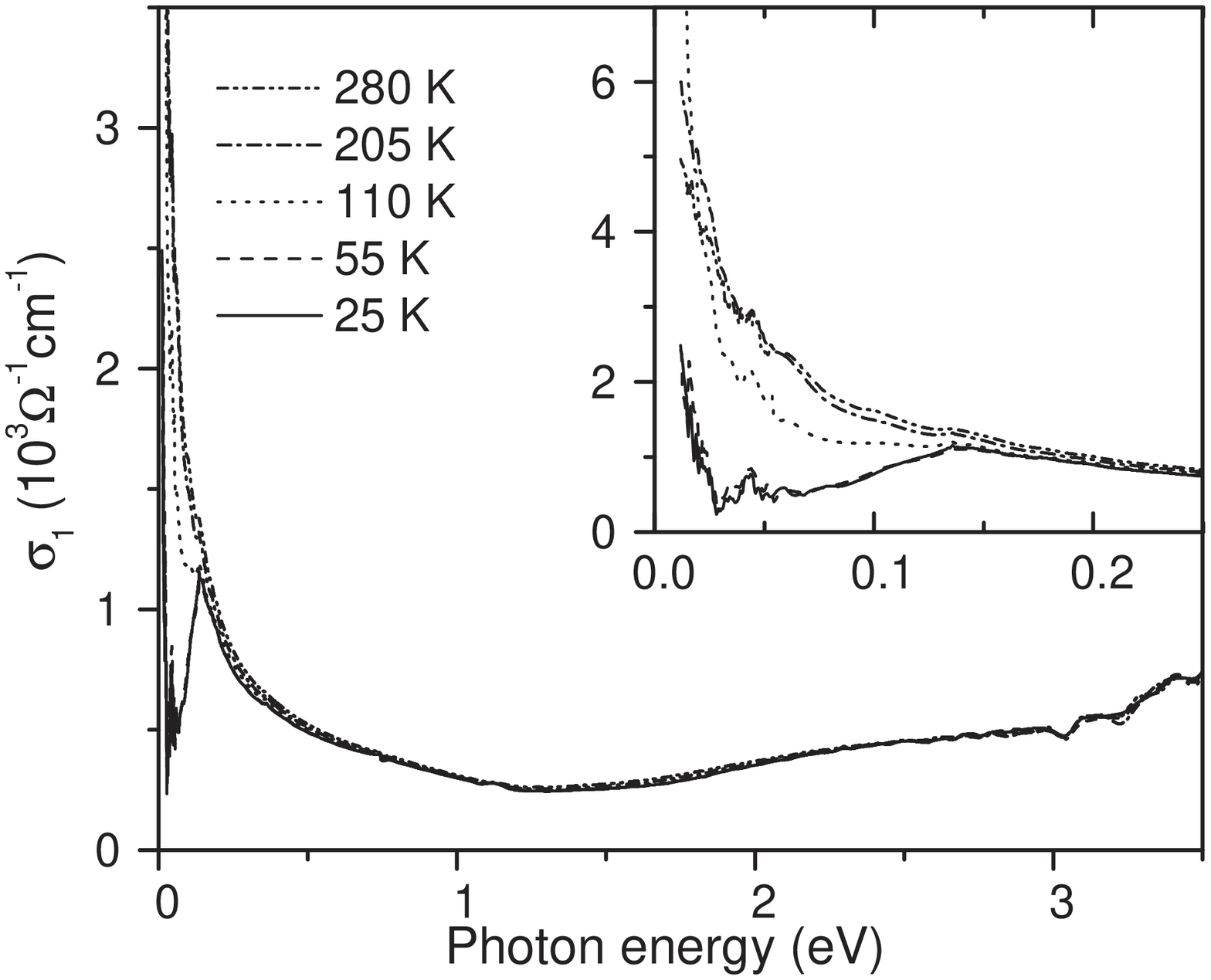}}
   \caption{In-plane optical conductivity of Bi2223 at selected
   temperatures. The inset displays the low energy part of the
   spectrum.}
   \label{fig1}
\end{figure}

In Fig. \ref{fig1} we show the optical conductivity at selected
temperatures. The spectral and temperature dependence of
$\sigma_{1}(\omega)$ of Bi2223 is very similar to the one of
Bi2212\cite{QuijadaPRB99,MolegraafScience02}, although the
conductivity of Bi2223 is slightly larger, likely due to a higher
volume density of the CuO$_{2}$ planes in the tri-layer compound.
The strongest changes as a function of temperature occur at low
frequencies. In the normal state the dominant trend is the
narrowing of the Drude peak. The onset of superconductivity is
marked by the opening of the superconducting gap which suppresses
$\sigma_{1}(\omega)$ below about 120-140 meV, slightly higher than
in Bi2212. Such a large scale is apparently caused by a large gap
value in Bi2223, which amounts up to 60 meV, as shown by
tunnelling measurements \cite{KuglerJMMM05}.

\begin{figure}[ht]
   \centerline{\includegraphics[width=8.5cm,clip=true]{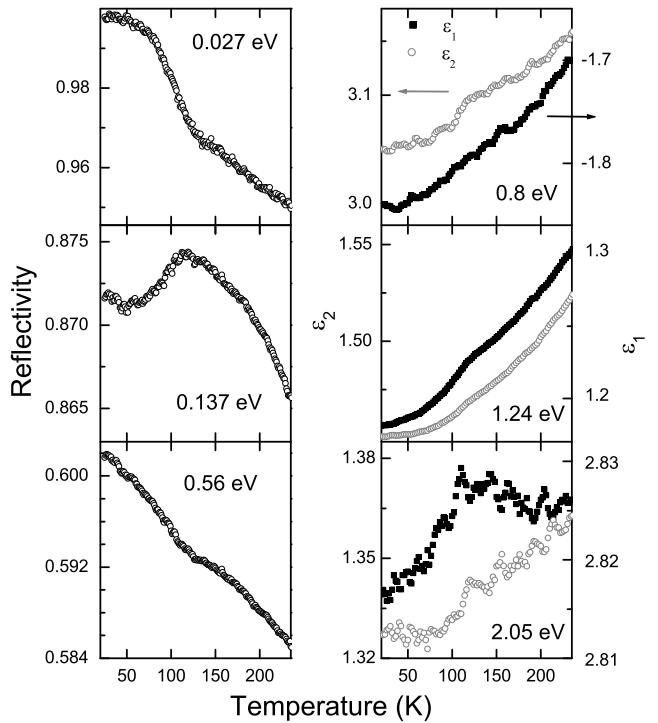}}
   \caption{Optical constants of Bi2223 at selected photon energies as a function of temperature.
   Left panel: normal incidence reflectivity at 0.027, 0.14 and 0.56 eV. Right panel: real and imaginary
   parts of the complex dielectric function at 0.8 1.24 2.05 eV. The photon energies are
   chosen close to the borders and in the middle of the experimental range.}
   \label{Repmeas}
\end{figure}

The much smaller absolute conductivity changes at higher energies,
which are not discernible at this scale, can be better seen in
Fig. \ref{Repmeas} where we show the temperature dependent optical
constants taken at selected photon energies. The change induced by
superconductivity in the optical constant is clearly visible as a
kink at $T_c$ for energies up to at least 2 eV, which tells that
the energy range where the redistribution of spectral weight takes
place is very large.

\section{Integrated spectral weight}

\subsection{Experimental determination of $W(\Omega_c)$}

The extraction of the spectral weight $W(\Omega_c)$ from the
measured spectra is a delicate issue. Formally, one has to integrate
the optical conductivity over a broad frequency range, including the
region below the low-frequency experimental cutoff $\Omega_{min}$
(in our case about 100 cm$^{-1}$) containing the condensate
$\delta$-peak (below $T_{c}$) at $\omega=0$ and a narrow
quasiparticle peak. According to a frequently occurring
misconception the existence of such a cutoff inhibits the
calculation of this integral. Indeed if only the real part of the
optical conductivity in some finite frequency interval was
available, clearly an essential piece of information needed to
calculate $W(\Omega_c)$ would be missing, namely $\sigma_1(\omega)$
below $\Omega_{min}$. However, due to the fact that the real and
imaginary part of the dielectric constant are related non-locally
via the Kramers-Kronig transformation, any change in one of them
will affect the other in a broad region of the spectrum. In
particular, any change of $\sigma_{1}(\omega)$ below $\Omega_{min}$
must influence $\epsilon_{1}(\omega)$ at higher frequencies. Since
the latter is measured independently (directly by the ellipsometry
above 0.75 eV cm$^{-1}$ and indirectly via the reflectivity in the
infrared), it puts constraints on the possible values of
$\sigma_{1}(\omega)$ below $\Omega_{min}$ and $W(\Omega_c)$.
Obviously, these constraints are going to be the more tight the more
accurately the optical constants are determined in the accessible
interval.

According to the KK relation
\begin{equation}
\epsilon_{1}(\omega) = 1+
8\wp\int_{0}^{\infty}\frac{\sigma_{1}(\omega^{\prime})d\omega^{\prime}}{\omega^{\prime
2}-\omega^{2}}
\end{equation}
\noindent the leading contribution of $\sigma_{1}(\omega)$ below
$\Omega_{min}$ to $\epsilon_{1}(\omega)$ above $\Omega_{min}$ is
proportional to the integral
$\int_{0}^{\Omega_{min}}\sigma_{1}(\omega)d\omega$ whereas
$\epsilon_{1}(\omega)$ is much less sensitive to the spectral
details of $\sigma_{1}(\omega)$ below
$\Omega_{min}$\cite{BozovicPRB90,KuzmenkoPRB05}. For example, the
contributions of the superfluid condensate and of a narrow
quasiparticle peak (provided that its width $\gamma
\ll\Omega_{min}$) to $\epsilon_{1}$ (and thus to the reflectivity)
at high frequencies are almost indistinguishable. Therefore the
value of the integral of $W(\Omega_c)$ can be model-independently
determined from our experimental data, while resolving the details
of $\sigma_{1}(\omega)$ below 100 cm$^{-1}$, for example the
separation of the superfluid density and quasiparticle spectral
weight, is not possible.

\begin{figure}[ht]
   \centerline{\includegraphics[width=6cm,clip=true]{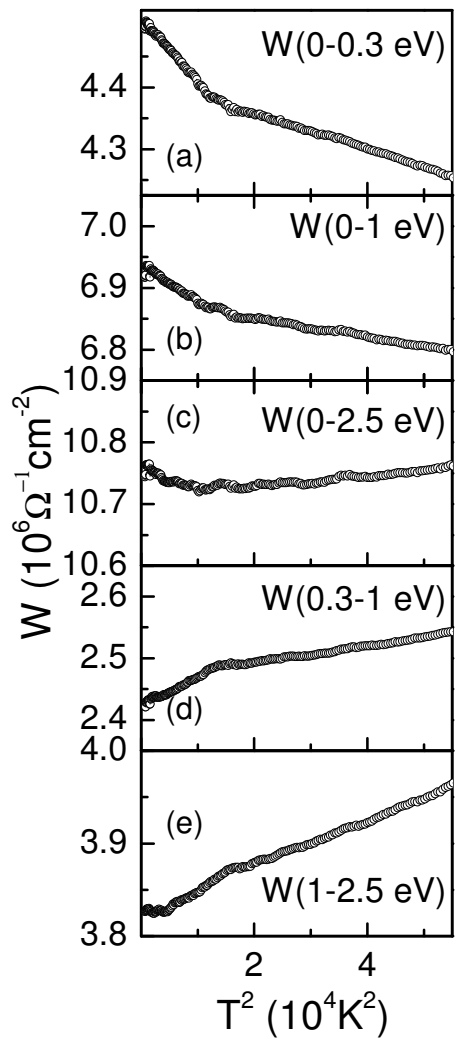}}
   \caption{Integrated spectral weight as a function of temperature (plotted vs. $T^{2}$
   for the reasons described in the text. The results corresponding to different
   ranges of integration are shown: (a) from 0 to 0.3 eV, (b) from 0 to 1 eV, (c) from 0 to 2.5,
   (d) from 0.3 to 1 (e) from 1 to 2.5 eV.}
   \label{SWtr1}
\end{figure}

In practice, the realization of the extra bounds on the value of
$W(\Omega_c)$ using the additional information contained in the
real part of the dielectric function can be done via the
aforementioned procedure of variational Kramers-Kronig constrained
fitting\cite{KuzmenkoRSI05}. Essentially, this is a modeling of
the data with a very large number of narrow oscillators, which are
added to the model dielectric function until all the fine details
of the measured spectra are reproduced. Importantly, the model
function always satisfies the KK relations. Once a satisfactory
fit of both reflectivity in the infrared region and
$\epsilon_{1}(\omega)$ and $\epsilon_{2}(\omega)$ above 0.75 eV is
obtained, we use the integral of $\sigma_{1}(\omega)$ generated
analytically by this multi-oscillator model as an estimate of the
spectral weight. Since the number of oscillators is very large and
their parameters are all automatically adjustable, this procedure
is essentially model independent. The spectral weight as a
function of temperature is shown in Fig. \ref{SWtr1}, for
different cut off frequencies.

\begin{figure}[ht]
   \centerline{\includegraphics[width=8cm,clip=true]{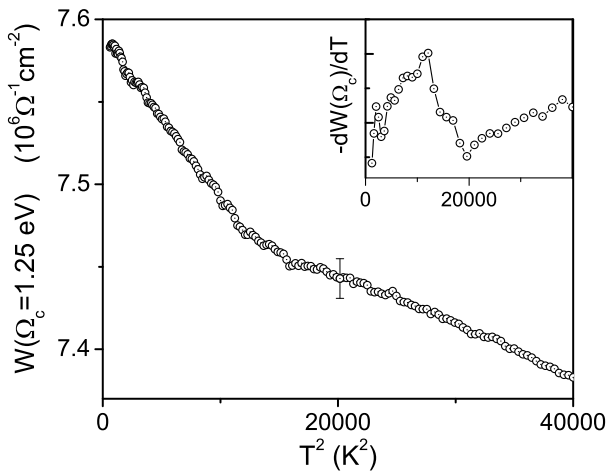}}
   \caption{Spectral weight $W(\Omega_{c},T)$ at cut of frequency of 1.25 eV plotted as a function of $T^2$. The errorbar is the maximum spread of the
   values of $W(\Omega_{c},T)$, obtained using three different low energy extrapolations shown in the panels of Fig.\ref{SWtr2}.}
   \label{SWtr22}
\end{figure}
\begin{figure}[ht]
   \centerline{\includegraphics[width=8cm,clip=true]{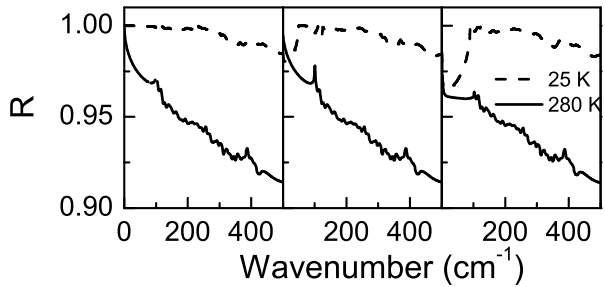}}
   \caption{Illustration of the numerical test described in the text: reflectivity curves down to zero frequency for different
   positions of the low energy peak, from left to right 0, 60 and 90 cm$^{-1}$.}
   \label{SWtr2}
\end{figure}

In order to illustrate that the measured spectra can indeed anchor
the value of the low frequency spectral weight without the need of
the low-frequency extrapolations of reflectivity (as it would be
the case in the conventional KK transform of reflectivity), we
performed the following numerical test. In the superconducting
state, the model contains a narrow oscillator centered at zero
frequency which accounts for the spectral weight of the condensate
and a narrow quasiparticle peak. We artificially displaced this
oscillator from the origin to a finite frequency, yet below
$\Omega_{min}$. The least mean square routine then readjusted all
the other parameters in order to produce a new best fit. After
that the overall fitting quality (the mean-squared error
$\chi^{2}$) and the low frequency spectral weight remained almost
unchanged with respect to the initial values, which is indicated
by the errorbar in Fig. \ref{SWtr22}, while the reflectivity below
100 cm$^{-1}$ now shows a very different behavior, as it is shown
in Fig. \ref{SWtr2}. This demonstrates that the details of
$\sigma_1(\omega)$ and $R(\omega)$ below 100 cm$^{-1}$ are not
essential for the determination of $W(\Omega_c$).

\subsection{Temperature dependent redistribution of spectral weight}

\begin{figure}[ht]
    \centerline{\includegraphics[width=8.5cm,clip=true]{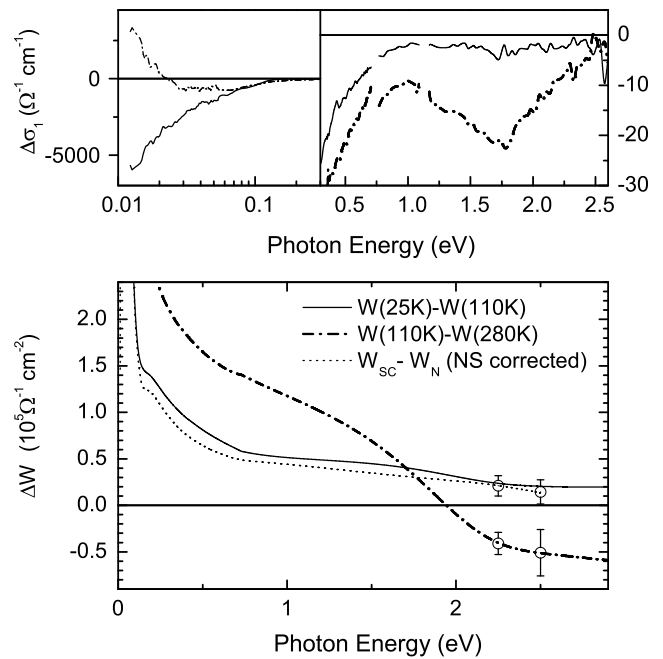}}
    \caption{Difference spectra of the in-plane optical conductivity and of the integrated spectral weight.
    Top panel: $\sigma_1(\omega,25 K)$-$\sigma_1(\omega, 110 K)$ (solid line) and
    $\sigma_1(\omega,110 K)$-$\sigma_1(\omega,280 K)$ (dashed-dotted line).
    Bottom panel: $W(\omega, 25K)$-$W(\omega, 110K)$ (solid line),
    $W(\omega,110K)$-$W(\omega,280K)$(dashed-dotted line) and $W_{SC}-W_{N}$, corrected for the normal
    state temperature dependence as descibed in the text (dotted line).}
    \label{delsig}
\end{figure}

It is important to establish the relevant cutoff energy in the
spectral weight integral. The optical conductivity of Bi2223 (see
Fig.\ref{fig1}) as well as the one of
Bi2212\cite{QuijadaPRB99,MolegraafScience02} shows a minimum
around 1 eV separating the Drude peak from the lowest-energy
interband transition, which is believed to be a charge transfer
excitation. It is interesting that such a separation is even more
pronounced if we plot the difference between conductivity spectra
at $T_{c}$ and a high temperature:
$\sigma_1(\omega,110K)-\sigma_1(\omega,280K)$ (see upper panel of
Fig. \ref{delsig}). A sharp upturn below 20 meV, a dip at around
50 meV and a slow approaching of this difference to zero (while
being negative) as the energy increases can be attributed to the
narrowing of the Drude peak \cite{KuzmenkoPRB05}. However, above
the same characteristic energy of about 1 eV this monotonic trend
totally disappears. Instead,
$\sigma_1(\omega,110K)-\sigma_1(\omega,280K)$ shows a dip at 1.5-2
eV, which corresponds to the removal of spectral weight from this
region, as the system is cooled down. The difference
$\sigma_1(\omega,25K)-\sigma_1(\omega,110K)$ between the two
spectra in the superconducting state also shows that the effect of
the narrowing of the Drude peak does not noticeably extend above 1
eV (Fig. \ref{delsig}).

We can learn more from the corresponding differences of the
integrated spectral weight $W(\omega,110K)-W(\omega,280K)$ and
$W(\omega,25K)-W(\omega,110K)$ which are displayed in the lower
panel of Fig. \ref{delsig}. In order to separate the effect of the
superconducting transition from the temperature dependence already
present in the normal state, we additionally plot the 'normal-state
corrected' spectral weight difference of the superconducting state
relative to the normal state $W_{SC}-W_N$ calculated according to
the procedure described in Section \ref{TempMod}. Not surprisingly,
all curves show an intense spectral structure below $\sim$ 0.3 eV as
a result of strong changes of the shape of the Drude peak with
temperature. However between 0.3-0.5 and 1.0-1.5 eV the spectral
variation is weak and, importantly, both normal- and
superconducting-state differences remain positive in this 'plateau'
region. This indicates an increase of the intraband spectral weight
as the sample is cooled down and an extra increase in the
superconducting state. Above 1.5-2 eV, the normal-state difference
$W(\omega,110K)-W(\omega,280K)$ decreases rapidly and becomes
negative, which suggests that spectral weight is transferred between
the charge-transfer and the intraband regions. In contrast, the
superconducting state differences (both normal-state corrected and
not) continue decreasing slowly and remain positive up to at least
2.5 eV, which means that the Ferrell-Glover-Tinkham sum rule is not
yet recovered at this energy.

Another way to visualize the spectral weight transfer is to plot
$W(\Omega_{c})$ as a function of temperature for different cutoff
energies. In Fig. \ref{SWtr1}(a-c) we present such curves for
$\Omega_{c}$ = 0.3, 1 and 2.5 eV.

One can immediately notice that the curves $W(0.3 eV, T)$ and $W(1
eV, T)$, apart from different absolute values, have almost
identical temperature dependencies; accordingly, the integrated
spectral weight between 0.3 and 1 eV (Fig. \ref{SWtr1}d) shows a
very little variation with temperature. This is, of course, a
manifestation of the existence of the discussed above 'plateau'
region in the frequency dependent spectral weight differences
(Fig. \ref{delsig}). This observation is in line with the
theoretical findings of Wrobel {\em et al}\cite{WrobelJPCM03} who
pointed out that spectral weight integrated to the hopping
parameter $t$ $\sim$ 0.3-0.4 eV is representative of the kinetic
energy of the t-J model.

Above $T_c$, the spectral weight $W(T)$ for the cutoff of 1 eV
increases gradually with cooling down in a virtually $T^2$
fashion, which is most clearly seen when $W(\Omega_{c})$ is
plotted versus $T^2$ (Fig. \ref{SWtr1}a,b). A similar normal state
behavior was observed in optimally and underdoped
Bi2212\cite{MolegraafScience02} and in
La$_{2}$CuO$_{4}$\cite{OrtolaniPRL05}. Although a $T^2$ term
follows trivially from the Sommerfeld expansion for the
temperature broadening of the Fermi-Dirac distribution, the
absolute value of this term turns out to be several times larger
than what one expects from this
expansion\cite{BenfattoPRB05,ToschiPRL05}. The DMFT calculations
within the Hubbard model \cite{ToschiPRL05} showed that this may
be caused by strong correlation effects. On the other hand, it was
recently pointed out\cite{KarakozovCM05,Benfatto06} that the
temperature dependence of the one-electron spectral function due
to inelastic electron-boson scattering contributes to the overall
temperature dependence of the optical sum rule much stronger than
the Sommerfeld term. The extra contribution though is
predominantly $T$-linear if the boson energy is
small\cite{Benfatto06}.

At the superconducting transition the curve $W(T)$ shows a sharp
{\em upward} kink (slope change) close to $T_c$ = 110 K. The same
effect was observed in optimally and underdoped Bi2212
\cite{MolegraafScience02}. In order to directly compare these data
to the results of Ref. \onlinecite{MolegraafScience02}, we plot in
Fig. \ref{SWtr22} the $W(\Omega_c,T)$ for $\Omega$=1.25 eV
together with the temperature derivative. One should stress that
the corresponding kinks are already observed on the temperature
dependence of directly measured optical constants (see
Fig.\ref{Repmeas}). By extrapolating the normal state trend to T =
0 K, we can estimate the size of the superconductivity induced
spectral weight transfer in the intraband region: $\Delta W
\approx 0.8\times10^5$ $\Omega^{-1}$cm$^{-2}$, which is about 1\%
of the total intraband spectral weight (as shown by
Fig.\ref{SWtr1}).

Remarkably, the upward kink of $W(\Omega_{c},T)$ at $T_c$ is still
observed for the cutoff of 2.5 eV (Fig. \ref{SWtr1}c) suggesting
that superconductivity induced spectral weight transfer involves
energies above the charge transfer gap, as we could already see
from Fig. \ref{delsig}. In the context of the Hubbard model, the
integrated spectral weight corresponds to the kinetic energy of
the Hubbard hamiltonian \cite{WrobelJPCM03} when the cut off
frequency is set much higher than the $U$. One would be tempted
conclude from the superconductivity-induced increase of W(2.5eV)
that also the total kinetic energy of the Hubbard model is lowered
below T$_c$, as observed in the cluster DMFT calculations
\cite{MaierPRL04}. We stress, however, that in order to make a
definitive statement about the kinetic energy in the Hubbard model
one should extend the cutoff frequency much higher than 2.5 eV.
Unfortunately, the noise level at higher energies precludes the
observation of the small superconductivity-induced effects.
Another problem is that this range overlaps with strong absorption
range due to charge transfer from oxygen to copper and that those
transitions may have temperature dependencies for a variety of
reasons which are not related to the kinetic energy lowering of
the charge carriers.

\subsection{Temperature modulation analysis at $T_{c}$}\label{TempMod}

The existence of relatively sharp kinks (slope discontinuities) at
$T_{c}$ on the curves of the temperature dependence of various
optical functions (see Fig.\ref{Repmeas}) enables an alternative
way to quantify the superconductivity-induced spectral weight
changes which gives perhaps a better feeling of the error bars
involved \cite{KuzmenkoPRL03}. We recently applied this procedure,
which is essentially a temperature-modulation method, to a similar
set of data on optimally doped Bi2212 \cite{KuzmenkoPRB05}.

\begin{figure}[ht]
    \centerline{\includegraphics[width=8.5cm,clip=true]{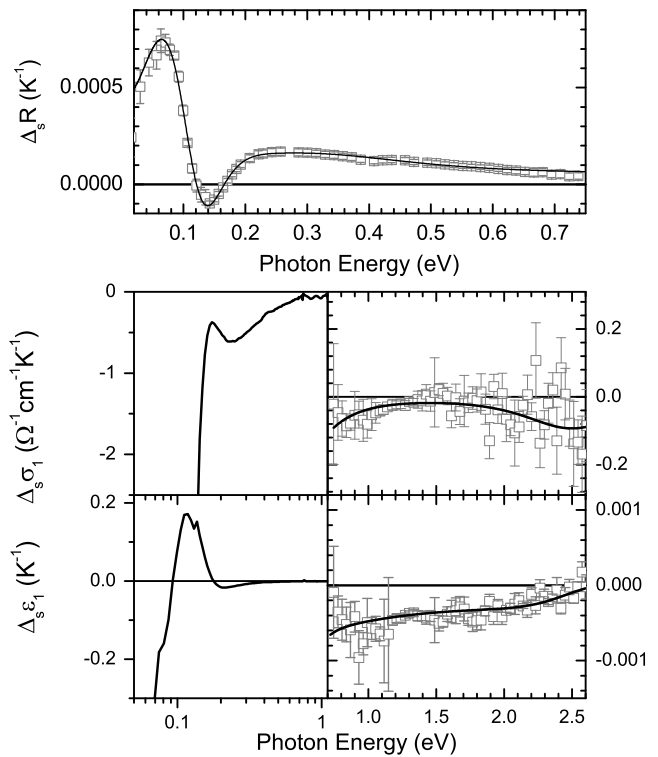}}
    \caption{Slope-differential spectra of the complex dielectric function,
    obtained experimentally (open symbols) together with the multi-oscillator fitting curves (solid
    lines) as described in the text.}
    \label{kinkan}
\end{figure}

In order to separate the superconductivity-induced effect from the
large temperature dependence in the normal state not related to
the onset of superconductivity we apply the 'slope-difference
operator' $\Delta_s$ defined as \cite{KuzmenkoPRL03}:
\begin{equation}\label{equation2}
        \Delta_s f(\omega) \equiv
        \left.\frac{\partial f(\omega,T)}{\partial T}\right|_{T_c + \delta} -
        \left.\frac{\partial f(\omega,T)}{\partial T}\right|_{T_c -
        \delta},
\end{equation}
\noindent where $f$ stands for any temperature dependent function.
In Fig.\ref{kinkan} the slope-difference spectra
$\Delta_{s}R(\omega)$, $\Delta_{s}\epsilon_{1}(\omega)$ and
$\Delta_{s}\sigma_{1}(\omega)$ are displayed with the error bars,
which we determined from the temperature dependent curves such as
shown in Fig.\ref{Repmeas}, using a numerical procedure, described
in Ref.\onlinecite{KuzmenkoPRB05}. Since $\Delta_s$ is a linear
operator, the KK relation between $\epsilon_{1}$ and
$\epsilon_{2}$ holds also for the slope-difference spectra
$\Delta_{s}\epsilon_{1}(\omega)$ and
$\Delta_{s}\epsilon_{2}(\omega)$. Thus we can fit the latter
spectra with a multi-oscillator Drude-Lorentz model, which
automatically satisfies the KK relations. If the number of
oscillators is large enough, the procedure becomes essentially
model-independent. Using the same dielectric function we can
additionally calculate the slope-difference spectrum of
reflectivity as it is related to $\Delta_{s}\epsilon_{1}(\omega)$
and $\Delta_{s}\epsilon_{2}(\omega)$:
\begin{eqnarray}
\Delta_{s}R(\omega)= \frac{\partial R}{\partial
\epsilon_{1}}(\omega,T_{c}) \Delta_{s}\epsilon_{1}(\omega)+
\frac{\partial R}{\partial \epsilon_{2}}(\omega,T_{c})
\Delta_{s}\epsilon_{2}(\omega).
\end{eqnarray}

\noindent The functions $\frac{\partial R}{\partial
\epsilon_{1}}(\omega,T_{c})$ and $\frac{\partial R}{\partial
\epsilon_{2}}(\omega,T_{c})$ can be derived from the analysis of
optical spectra at $T_{c}$\cite{KuzmenkoPRB05}.

The best fit of the slope-difference optical constants for Bi2223
is shown by the solid curves in Fig.\ref{kinkan}. From this we
also calculate the slope-difference integrated spectral weight
$\Delta_{s} W(\omega) = \int_0^{\omega} \Delta_{s}
\sigma_1(\omega^\prime) d\omega^\prime$ as shown in
Fig.\ref{kinkan}(a). At $\Omega_{c}$ = 1 eV we obtain the value of
$\Delta_{s}W$ $\approx$ +1100 $\Omega^{-1}$cm$^{-2}$K$^{-1}$ which
corresponds to the superconductivity-induced increase of spectral
weight, in agreement with the previous analysis.

\begin{figure}[ht]
    \centerline{\includegraphics[width=8.5cm,clip=true]{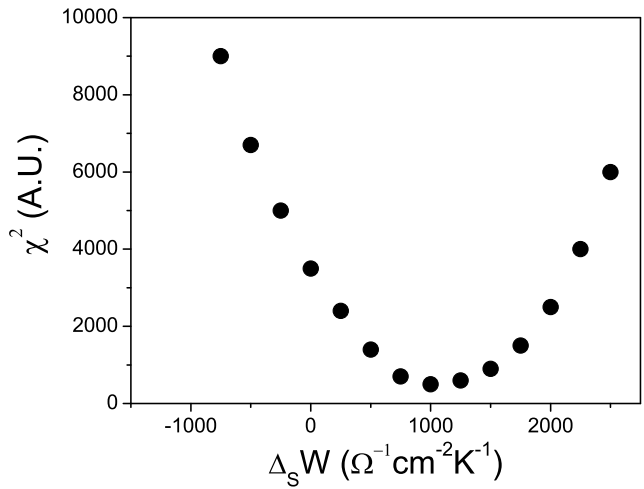}}
    \caption{The demonstration that $\Delta_{s}W(\Omega_{c})$ is
    well defined by the available data ($\Omega_{c}$=1 eV). The mean squared error
    $\chi^2$ for the total fit of $\Delta_{s}R(\omega)$, $\Delta_{s}\epsilon_{1}(\omega)$
    and $\Delta_{s}\sigma_{1}(\omega)$ (shown in
    Fig.\ref{kinkan})
    as a function of $\Delta_{s}W(\Omega_{c})_{imposed}$. More description is given in the text.}
    \label{chisq}
\end{figure}

To check that the value of $\Delta_{s}W$ is well defined by the
present set of experimental spectra we repeated the fitting
routine while forcing $\Delta_{s}W(\Omega_{c}=1 eV)$ to be equal
to some imposed value $\Delta_{s}W(\Omega_{c})_{imposed}$. We did
this for different values of $\Delta_{s}W(\Omega_{c})_{imposed}$
and monitored the quality of the fit, as expressed by the
mean-squared error $\chi^2$. In Fig. \ref{chisq} we plot $\chi^2$
as a function of $\Delta_{s}W(\Omega_{c})_{imposed}$. The best fit
quality is, of course, achieved for the mentioned value $\Delta_S
W(\Omega_c)_{imposed} \approx 1100$
$\Omega^{-1}$cm$^{-2}$K$^{-1}$. It is evident that the fit quality
deteriorates rapidly as $\Delta_S W(\Omega_c)_{imposed}$ is
dragged away from this value. For example, the value of $\chi^{2}$
for the case $\Delta_S W(\Omega)_{imposed}$ = 0 (which would be
the full recovery of the sum rule at 1 eV) is about 10 times
larger than the optimal value; the corresponding data fit should
be regarded as unacceptable. As it was discussed in
Ref.\onlinecite{KuzmenkoPRB05} and in this paper, this is due to
the fact that any change of the low-frequency spectral weight
inevitably affects the value of $\epsilon_{1}(\omega)$ and
$R(\omega)$ at higher frequencies.

To compare the approach described in this Section and the one from
previous Section, we derive from the curve $W(T)$, plotted in Fig.
\ref{SWtr1} $\Delta_{s}W$($\Omega_{c}$=1 eV)$\approx$ 1500
$\Omega^{-1}$cm$^{-2}$K$^{-1}$. This is not far from the
aforementioned value of 1100 $\Omega^{-1}$cm$^{-2}$K$^{-1}$ which
quantitatively confirms the central assertion of this paper,
namely that the low frequency spectral weight increases below
$T_{c}$.


As it was discussed in Ref.\onlinecite{KuzmenkoPRB05}, the
'normal-state corrected' spectral weight difference of the
superconducting state relative to the normal state, $W_{SC}-W_N$
can be estimated from the slope-difference conductivity spectra:
\begin{equation}
W_{SC}-W_N\approx\alpha T_c \Delta_s W ,
\end{equation}

\noindent where $\alpha$ is a dimensionless coefficient. The
choice of $\alpha$ is suggested by the temperature dependence of
$W(\Omega_c)$\cite{KuzmenkoPRB05}. Since we observe in both normal
and superconducting state a temperature dependence close to $T^2$
then we choose $\alpha=1/2$. The corresponding curve $W_{SC}-W_N$
is shown as a dotted line in Fig.\ref{delsig}. One can see that it
is slightly smaller than the direct difference
$W(\omega,25K)-W(\omega,T=110K)$. The latter fact is not
surprising since the spectral weight in the normal state is
increasing as a function of temperature.

\section{Discussion and conclusions}

The superconductivity-induced increase of low-frequency spectral
weight, which implies the opposite to the BCS type lowering of the
electronic kinetic energy, was previously observed in the
under-doped and optimally doped bi-layer \bb\
\cite{MolegraafScience02,SantanderEPL03,KuzmenkoPRB05}. Recent
studies\cite{DeutscherPRB05,CarboneDopDep} show that this effect
changes sign for strongly overdoped samples of the Bi2212,
following the trend of the BCS model. A temperature dependence of
the low energy spectral weight in accordance to BCS theory has
also been claimed for optimally doped
YBa$_{2}$Cu$_{3}$O$_{7-\delta}$ and slightly underdoped
Bi2212\cite{BorisScience04}, although the latter result is
controversial\cite{KuzmenkoPRB05,SantanderCM05}. Hence the
'unconventional' superconductivity induced increase of low energy
spectral weight appears to be a property characteristic of the
Bi-based multi-layer cuprates at and below optimal doping, but not
of the entire cuprate family for all doping levels.

Recently an intriguing connection has been pointed out between the
superconductivity induced increase of $W(T)$ on the one hand, and
the drop of scattering rate on the other hand
\cite{MarsiglioPRB06}. Since the former involves a spectral weight
integral over 1 eV, whereas the latter is measured at microwave
frequencies, these two experimental observations are seemingly
unrelated. However, decreasing the scattering results in
sharpening of the occupation distribution in $k$-space. Hence a
decrease of scattering automatically implies a decrease of the
average kinetic energy, which in turn is observed as an increased
$W(T)$ when the sample turns from normal to superconducting. Since
the standard BCS model (without a change in scattering) already
predicts a superconductivity induced increase of the kinetic
energy, the net result of both effects (scattering change and
intrinsic BCS effect) will depend on the relative magnitude of
these two effects. This explanation successfully relates two
classes of experiments, without directly relating either one of
these experiments to a particular pairing mechanism. The starting
assumption of an anomalous drop in scattering is at this stage a
phenomenological one, and still requires a microscopic
explanation.

We observe that the increase of spectral weight below $T_{c}$ in
Bi2223 is larger than in Bi2212\cite{MolegraafScience02}. We
believe that this can be quite generally understood. It is well
known that in the BCS theory
\begin{equation}\label{BCSscaling}
\Delta E_{kin}\sim\Delta_{SC}^2\sim T_{c}^{2}.
\end{equation}
The first equality holds generally for situations where the
electronic occupation numbers $\langle n_{k}\rangle$ are
redistributed in an energy range $\Delta$ around $E_F$, whereas
the second is suggested simply by a dimensional analysis. Recent
STM studies \cite{KuglerJMMM05} evidence that the SC gap is indeed
larger in Bi2223 than in Bi2212 at a similar doping level.
Assuming that Eq. (\ref{BCSscaling}) is also valid in Bi2212 and
Bi2223 at optimal doping, and Eq. (\ref{equation1}) is exact, we
obtain $\Delta W_{Bi2223}/\Delta
W_{Bi2212}=T_{c,Bi2223}^{2}/T_{c,Bi2212}^{2}\approx1.6$. This
ratio value matches the experimental observation, supporting the
idea that the spectral weight transfer is intimately related to
the SC-induced redistribution of the occupation numbers.

In summary, we have observed a superconductivity-induced increase
of the in-plane low-frequency optical spectral weight
$W(\Omega_{c})$ in tri-layer \bbb\ near optimal doping from the
reflectivity and ellipsometric measurements. Comparison to the
case of optimally doped Bi2212 suggests that the size of the
spectral weight transfer scales with the value of the critical
temperature. In the normal state, the temperature dependence of
the $W(\Omega_{c})$ is essentially proportional to $T^{2}$.

\section*{ACKNOWLEDGMENTS}

We are grateful to F. Marsiglio, T. Timusk, N. Bontemps, A.F.
Santander-Syro, J. Orenstein, and C. Bernhard for stimulating
discussions and A. Piriou for technical assistance. This work was
supported by the Swiss National Science Foundation through the
National Center of Competence in Research "Materials with Novel
Electronic Properties-MaNEP".

\section{APPENDIX}

\subsection{Determination of the c-axis dielectric function}

In order to properly convert the pseudodielectric function
measured ellipsometrically on the ab-crystal surface to the true
dielectric function along the ab-plane, we additionally measured
the c-axis dielectric function using another crystal of Bi2223
grown under the same conditions.

\begin{figure}[ht]
   \centerline{\includegraphics[width=8cm,clip=true]{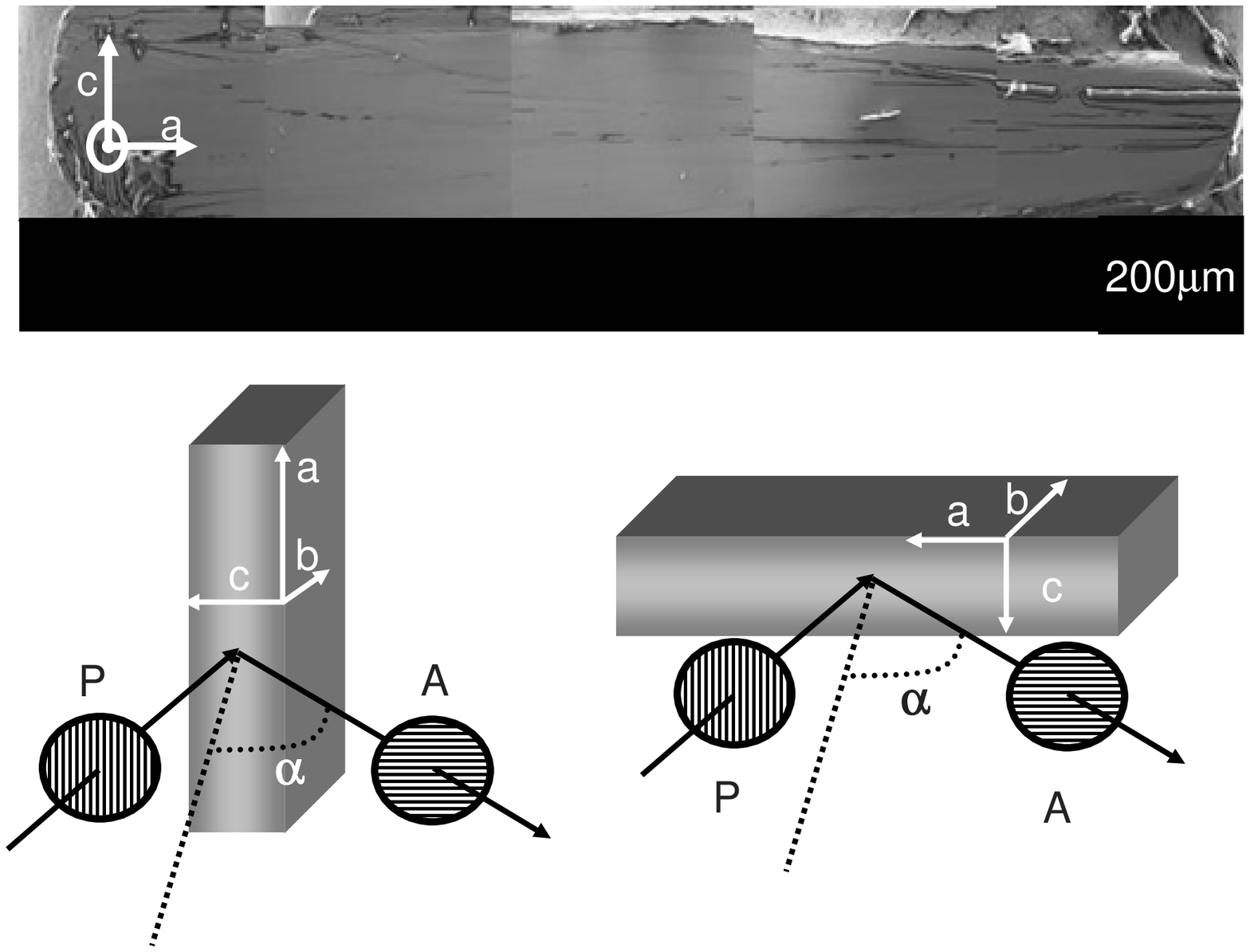}}
     \caption{Top: the image of the ac plane after polishing. Bottom: two geometries of the ellipsometry experiment.}
\label{geom}
 \end{figure}

We did spectroscopic ellipsometry from 6000 and 36000 cm$^{-1}$ on
an ac surface of dimensions ($a\times c$) 3$\times$0.8 mm$^2$
which we cut and polished with a diamond paper of 0.1 $\mu$m grain
size. The surface image is shown in Fig. \ref{geom}. Two
orthogonal orientations of the sample were used, designated as
(ac) and (ca), as shown in Fig.\ref{geom} which provided four
ellipsometric parameters
$\psi_{ac},\Delta_{ac},\psi_{ca},\Delta_{ca}$ at each frequency.
Assuming that $\epsilon_a \approx \epsilon_b$ we inverted four
corresponding expressions based on the Fresnel equations in order
to obtain four unknown variables $\epsilon_1^{ab}$,
$\epsilon_2^{ab}$, $\epsilon_1^c$ $\epsilon_2^c$. We applied this
procedure for three angles of incidence 65$^{\circ}$, 70$^{\circ}$
and 75$^{\circ}$ simultaneously in order to improve the accuracy
of the inversion.

We also measured the normal incidence reflectivity between 450 and
8000 cm$^{-1}$ on the same crystal plane for the electric field
parallel to the c-axis. This measurement agrees well with the
ellipsometrically determined $\epsilon_{c}(\omega)$, which
confirms the validity of the described inversion procedure.
Finally, the reflectivity and ellipsometry output were all fitted
simultaneously with a variational KK consistent function
\cite{KuzmenkoRSI05} in order to extend the frequency dependence
of $\epsilon_{c}(\omega)$ down to 450 cm$^{-1}$ while anchoring
the unknown phase of the reflectivity by the ellipsometric data.

The c-axis reflectivity $R_{c}(\omega)$, optical conductivity
$\sigma_{1c}(\omega)$ and the real part of the dielectric function
$\epsilon_{1c}(\omega)$ are displayed in Fig. \ref{casp}. We
observe a weak wavelength dependence for $\epsilon_{1c}(\omega)$,
which is in a good agreement with with a previous report by Petit
\emph{et al} \cite{PetitEPJB02}. However, we found a rather
different $\sigma_{1c}(\omega)$ which is likely due to the fact in
that in Ref.\onlinecite{PetitEPJB02} the reflectivity was measured
on a textured polycrystalline sample of Bi2223 and the
conductivity was obtained by a conventional Kramers-Kronig
transform. As it was pointed in Ref.\onlinecite{PetitEPJB02}, the
misalignment of the grains of the oriented ceramic can have a
considerable impact on the final result. The main advance compared
to these earlier results is that our samples were single crystals,
and we determined the real and imaginary part of the dielectric
tensor in a direct way using ellipsometry, without the need of a
Kramers-Kronig transformation.

\subsection{The sensitivity of the in-plane dielectric function
to the c-axis correction}

\begin{figure}[ht]
   \centerline{\includegraphics[width=9cm,clip=true]{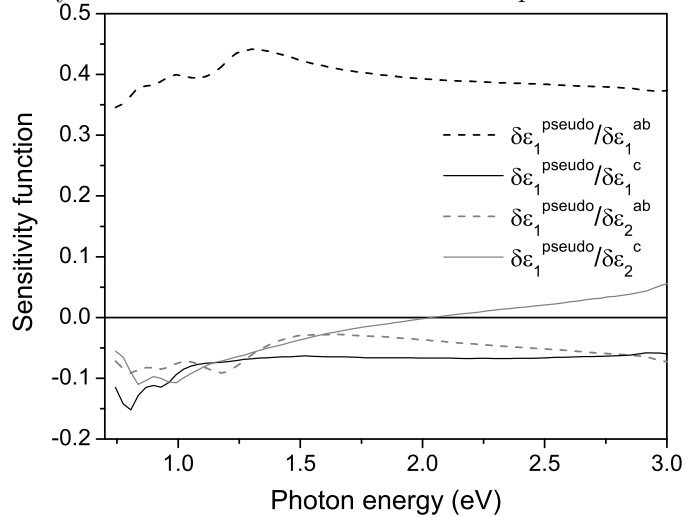}}
     \caption{Real and imaginary parts of the sensitivity functions
     $\partial\epsilon_{pseudo}/\partial\epsilon_{ab}$ and $\partial\epsilon_{pseudo}/\partial\epsilon_{c}$
     defined in the text.}
\label{f1}
 \end{figure}

The pseudodielectric function measured on the ab oriented crystal
surface depends on both components of the dielectric tensor and
the angle of incidence $\theta$: $\epsilon_{pseudo} =
f(\epsilon_{ab}, \epsilon_{c},\theta)$. It was shown by
Aspnes\cite{AspnesJOSA80} that in this case the pseudodielectric
function should be much more sensitive to the ab-plane component,
which lies along the crossing line of the plane of incidence and
the sample surface, than to the c-axis one. In order to verify
that this is the case, we show in Fig. \ref{f1} the "sensitivity
functions" $\partial\epsilon_{pseudo}/\partial\epsilon_{ab}$ and
$\equiv\partial\epsilon_{pseudo}/\partial\epsilon_{c}$ calculated
for the actual angle of incidence (in our case 74$^{\circ}$) on
the base of described above ac-plane ellipsometry results.

\begin{figure}[ht]
   \centerline{\includegraphics[width=9cm,clip=true]{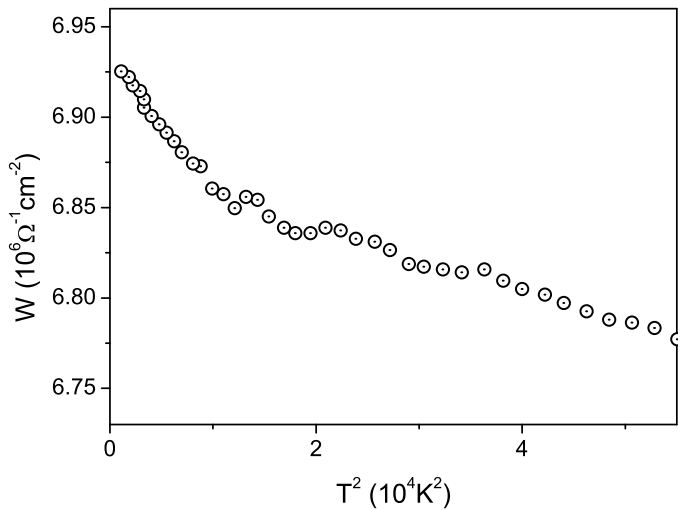}}
     \caption{Spectral weight as a function of temperature (for the cutoff of 1 eV) obtained
     when a temperature dependent c-axis data were used to correct the ab-plane pseudodielectric function for the admixture
     of the c-axis component. The temperature resolution
     in this plot is lower than in Fig.\ref{SWtr1} because the c-axis temperature dependent data have a resolution of 5
     K.}
\label{swc}
\end{figure}

One can see that the pseudodielectric function is indeed much less
sensitive to the c-axis component since
$\partial\epsilon_{pseudo}/\partial\epsilon_{c}$ is about 4-5
times smaller than
$\partial\epsilon_{pseudo}/\partial\epsilon_{ab}$. For this
reason, the temperature dependence of the c-axis dielectric
function is expected to have only a minor effect on that of the
measured pseudodielectric function. In order to verify this, we
performed the c-axis correction of the ab-plane pseudodielectric
function and calculated the spectral weight integral
$W(\Omega_{c}=1eV$) using in the first case the temperature
dependent c-axis dielectric function and in the second case a
constant, temperature-averaged one. The resulting temperature
dependence of $W(\Omega_{c}$) in the former case is shown in Fig.
\ref{swc}, while the one in the latter case is shown in
Fig.\ref{SWtr1}. One can see that accounting for the temperature
dependence of the c-axis dielectric function does not have any
significant influence on temperature dependence of
$W(\Omega_{c}$), except for a stronger scatter of the datapoints
as a result of the inevitable noise introduced by extra
measurement on a small crystal surface. Therefore we used the
temperature averaged c-axis data to correct the ab-plane
pseudodielectric function in the main part of this paper.

\end{document}